\documentclass[]{emulateapj}
\def\pppm{\rm P^3M}
\def\mpchi{\,h^{-1}{\rm {Mpc}}}
\def\mpchii{\,h{\rm {Mpc}^{-1}}}

\def\msun{\,h^{-1}{\rm M_\odot}}
\def\k{\mathbf{k}}
\def\x{\mathbf{x}}

\begin{document}

\title{A determination of dark matter bispectrum with a large set of
$N$-body simulations}

\author{H. Guo\altaffilmark{1,2}, Y. P. Jing\altaffilmark{1}}
\altaffiltext{1}{Key Laboratory for Research in Galaxies and
Cosmology, Shanghai Astronomical Observatory, Chinese Academy of
Sciences, Nandan Road 80, Shanghai 200030, China; guoh@shao.ac.cn,
ypjing@shao.ac.cn.} \altaffiltext{2}{Graduate School of the Chinese
Academy of Sciences, 19A, Yuquan Road, Beijing, China}

\begin{abstract}

We use a set of numerical $N$-body simulations to study the
large-scale behavior of the reduced bispectrum of dark matter and
compare the results with the second-order perturbation theory and
the halo models for different halo mass functions. We find that the
second-order perturbation theory (PT2) agrees with the simulations
fairly well on large scales of $k<0.05\mpchii$, but it shows a
signature of deviation as the scale goes down. Even on the largest
scale where the bispectrum can be measured reasonably well in our
simulations, the inconsistency between PT2 and the simulations
appears for the colinear triangle shapes. For the halo model, we
find that it can only serve as a qualitative method to help study
the behavior of $Q$ on large scales and also on relatively small
scales. The failure of second-order perturbation theory will also
affect the precise determination of the halo models, since they are
connected through the 3-halo term in the halo model. The 2-halo term
has too much contribution on the large scales, which is the main
reason for the halo model to overpredict the bispectrum on the large
scales. Since neither of the models can provide a satisfying
description for the bispectrum on scales $k\sim 0.1 \mpchii$ for the
requirement of precision cosmology, we release the reduced
bispectrum of dark matter on a large range of scales for future
analytical modeling of the bispectrum.

\end{abstract}

\keywords{gravitational lensing---dark matter--- cosmology:
theory--- galaxies: formation}

\section{Introduction}

The understanding of the formation, clustering, and evolution of
galaxies needs to involve the dark matter in modern cosmological
models. In the current standard paradigm, i.e., $\Lambda$ cold dark
matter (CDM) models, dark matter has negligible velocity in the
early universe and dominates the matter density in the present
universe. Starting from an initially tiny Gaussian density
fluctuation, the density distribution of CDM would keep its
Gaussianity throughout the whole evolution in the linear-order
perturbation (e.g., \citealt{Peebles80,Bernardeau02,Sefusatti07}).
However, the dynamical evolution of the fluctuations, through the
mode coupling of different scales, would result in non-Gaussianity
of density distribution on all scales, including the conventional
linear regime. Various statistical tools are used in the literature
to describe the density distribution, and can be further
incorporated into the research of non-Gaussianity. For example, the
two-point correlation function(2PCF) $\xi(r)$ is employed to
describe the correlation of two objects at a separation $r$, while
the three-point correlation function(3PCF) is used to describe the
next-order statistics in the series of connected $N$-point
correlation functions(NPCFs) \citep{Peebles80}. In the linear-order
perturbation theory, a Gaussian field is completely determined by
the 2PCF. Therefore, the 3PCF is the lowest-order non-Gaussian
quantity in the series, which can give us important clues to the
properties of nonlinear evolution.

The dark matter 3PCF and its Fourier-space counterpart, the
bispectrum, have been studied intensively over the years(e.g.,
\citealt{Fry84,Scoccimarro98,Hou05,Nichol06}). The theoretical work
mainly comprises two methods: the nonlinear perturbation
theory(hereafter PT; e.g., \citealt{Bernardeau02}, and the
references therein) and the halo model(hereafter HM;
\citealt{Jing98,Ma00,Peacock00,Seljak00,Scoccimarro01,Takada03,
Yang03,Wang04,Smith08}). Both methods have been applied to and
compared with simulations over a range of scales(e.g.,
\citealt{Scoccimarro01,Hou05,Smith08}). In addition, as a prior
assumption, PT is believed to be applicable on large scales where
the density fluctuation is small enough for a perturbative method to
be useful. It starts to fail as the scale goes down to the strongly
nonlinear regime.

The halo model approach, however, is based upon dark matter halos.
By utilizing halo correlation functions and the dark matter
distribution in the halos, the halo model can then generate the dark
matter clustering statistics without any consideration of the
scales.  An advantage of the halo model method is that after being
combined with the models of galaxy distribution in the host halos,
which is often called halo occupation distribution(HOD; e.g.,
\citealt{jmb98,Berlind02,Yang03,Zehavi04,Zheng04,Zheng05}), it can
produce the overall distribution of galaxies, which can be directly
compared with the observations of large-scale galaxy surveys. So,
the halo model method can also be used to predict the galaxy-biasing
relations(e.g., \citealt{Seljak00,Scoccimarro01,Verde02}).

The development of numerical simulations in recent years helps test
the validity and range of applicability of these theories(e.g.,
\citealt{Scoccimarro01,Fosalba05,Smith08}). On large scales, both PT
and HM are believed to be in good agreement with the simulations. On
smaller scales(quasi-linear and nonlinear regimes), PT breaks down
fairly quickly, while HM may still be a good approximation. However,
most of the previous simulation studies in the literature only
measure the bispectrum on the scales smaller than that of $k \sim
0.1\mpchii$, due to the limited dynamical range of their
simulations.

In this work, we use a set of high-resolution large-box cosmological
$N$-body simulations to accurately determine the dark matter
bispectrum on both large and small scales. The results will be
compared with the predictions of PT and HM in order to test their
validity in describing the bispectrum. It is important to determine
where and how well the theories are valid, since a determination of
the galaxy bispectrum in observation can yield a determination of
the galaxy bias factors on large scales if the bispectrum of dark
matter can be predicted \citep[e.g.,][]{Fry84,Fry93}. As we will see
that neither the second-order PT (hereafter PT2; without loop
corrections) nor the HM can provide a satisfying description for the
dark matter bispectrum on scales of $k\sim 0.1 \mpchii$ in the
standard of precision cosmology, we make the bispectrum of dark
matter on a large range of scales available to interested readers,
who want to use the data to construct a more accurate model for the
bispectrum.

This paper is organized as follows. In Section \ref{methods}, we
describe some basic formulae and the method used in this paper. We
introduce our simulations in Section \ref{simulations}. We compare
the simulations with PT and halo models in Section \ref{pt} and
Section \ref{hm}, respectively. We summarize our results in Section
\ref{conclusions}.

\section{Methods}\label{methods}

In a statistically homogeneous and isotropic universe, the density
fluctuation of the matter distribution is defined as
$\delta(\mathbf{x})=(\rho(\x)-\bar{\rho})/\bar{\rho}$, with the mean
density being $\bar{\rho}=\langle\rho(\x)\rangle$. Due to the
convenience in calculating bispectrum in Fourier space, we transform
$\delta(\x)$ to

\begin{equation}
\delta_{\k} = \int d^3 x \exp(-i \k \cdot \x) \delta(\x).
\end{equation}

By definition $\langle \delta_{\k} \rangle=0$. So the next two
nonvanishing terms in the series will be

\begin{eqnarray}
\langle \delta_{\k_1} \delta_{\k_2} \rangle &=&
\delta^\mathrm{D}(\k_1+\k_2) P(k_1)\\
\langle \delta_{\k_1} \delta_{\k_2} \delta_{\k_3} \rangle &=&
\delta^\mathrm{D}(\k_1+\k_2+\k_3) B(k_1,k_2,k_3),
\end{eqnarray}
where $\delta^\mathrm{D}(\k)$ is the three-dimensional Dirac delta
function and $P(k)$ and $B(k_1,k_2,k_3)$ are the Fourier transform
of the 2PCF and 3PCF, known as the power spectrum and bispectrum,
respectively. If we extend the perturbation theory to the second
order, it gives \citep{Fry84,Bernardeau02}

\begin{eqnarray}
B_{PT}(k_1,k_2,k_3) &=& F(\mathbf{k_1},\mathbf{k_2})
P_L(k_1)P_L(k_2)+ cyc, \label{eqn:Bk}
\\
F(\mathbf{k_1},\mathbf{k_2}) &=& (1 + \mu) + \frac{\mathbf{k_1}
\cdot \mathbf{k_2}}{k_1k_2}\left(\frac{k_1}{k_2} +
\frac{k_2}{k_1}\right)
\nonumber\\
&&+ (1 - \mu)\left(\frac{\mathbf{k_1} \cdot
\mathbf{k_2}}{k_1k_2}\right)^2,
\end{eqnarray}
where $P_L(k)$ is the linear power spectrum and
$\mu=3\Omega_m^{-2/63}/7$ denotes the weak dependence on cosmology.

It is often convenient to define the reduced bispectrum,

\begin{equation}
Q(k_1,k_2,k_3)=\frac{B_{PT}(k_1,k_2,k_3)}{P_L(k_1)P_L(k_2)+cyc},\label{eqn:qbp}
\end{equation}
where $k_1$, $k_2$, and $k_3$ set the three sides of the triangle
shape. Two sets of variables are frequently used for $Q$. We define
the three variables to be $k=k_1$, $u=k_2/k_1$, and
$v=(k_3-k_2)/k_1$, where $k_1 \leqslant k_2 \leqslant k_3$, $u \geq
1$, and $0 \leq v<1$. Therefore, $k$ sets the scale of the triangle,
while $u$ and $v$ determine its shape. In addition, if we change $v$
to $\alpha=arccos(\k_1 \cdot \k_2)/k_1 k_2$, ($0<\alpha<\pi$), we
get the other form as $Q(k,u,\alpha)$, which clearly shows the
relation between $k_1$ and $k_2$. Also, the shape dependence is
obviously described by $u$ and $\alpha$.  But this parametrization
has a disadvantage that when considering triangles of different
scales and different shapes together, we may have counted the set of
$(k_1,k_2,k_3)$ repeatedly, because in this case we only have
$0<k_3<k_1+k_2$. In order to avoid repeated counting, we still need
to impose a constraint of $k_3 \geq k_2$ when we count independent
triangles. We will use $Q(k,u,\alpha)$ as the preferred
parametrization when investigating the scale and shape dependence of
$Q$ in the following sections.

Although $Q(k,u,\alpha)$ is generally dependent on the scale and
shape of a triangle, in the second-order perturbation theory, if we
assume the power spectrum to be a power law, $Q$ is then completely
independent on the scale $k$. For the equilateral triangle shapes
$k_1=k_2=k_3$, from Equations (\ref{eqn:Bk}) and (\ref{eqn:qbp}) we
have $Q_{eq}=1/4+3\mu/4$, which is exactly constant. These important
properties of the reduced bispectrum $Q$ make it a better quantity
than the bispectrum $B(k_1,k_2,k_3)$ to characterize the
non-Gaussianity of the matter distribution. The shape dependence of
$Q$ reflects the influence of the gravitational instability, so it
can be significantly affected by the existence of large-scale
structures such as filaments \citep{Sefusatti05} for small
simulation volumes. In order to reduce such finite-volume effects,
large-box simulations are necessary to compute the bispectrum as we
do in this paper.

\section{$N$-body Simulations}
\label{simulations}

The cosmological model considered here is a canonical spatially flat
CDM model with the density parameter $\Omega_m=0.268$, the
cosmological constant $\Omega_\Lambda=0.732$, the Hubble constant
$h=0.71$, and the baryon density parameter $\Omega_b=0.045$. The
primordial density field is assumed to be Gaussian with a
scale-invariant power spectrum $\propto$$k$. For the linear
spectrum, we generate it from the CMBfast code \citep{Seljak96}, and
the normalization is set by $\sigma_8=0.85$, where $\sigma_8$ is the
present linear rms density fluctuation within a sphere of radius
$8\mpchi$.

\begin{deluxetable}{cccccc}
\tablewidth{0pt} \tablecolumns{4} \tablecaption{Simulation
parameters} \tablehead{\colhead{boxsize($\mpchi$)} & \colhead
{particles} & \colhead {realizations} & \colhead {$m_{\rm
particle}$}} \startdata
600 & $1024^3$ & 3 & $1.5\times10^{10}h^{-1}M_\odot$\\
1200 & $1024^3$ & 4 & $1.2\times10^{11}h^{-1}M_\odot$\\
1800 & $1024^3$ & 4 & \ $4.0\times10^{11}h^{-1}M_\odot$
\enddata
\label{tab:simu}
\end{deluxetable}

We use an upgraded version of the Particle-Particle-Particle-Mesh
($\pppm$) code of \citet{js98,js02} to simulate structure formation
in the universe. The code has now incorporated the multiple level
$\pppm$ gravity solver for high-density regions \citep{js00}. In
order to have a large mass resolution range, we run a total of $11$
simulations with $1024^3$ particles in different simulation boxes,
which we hereafter denote by $L600$, $L1200$, and $L1800$ by
different box sizes(Table \ref{tab:simu}) \citep{Jing07}. The
simulations were run on an SGI Altix 350 with 16 processors with
OPENMP parallelization in Shanghai Astronomical Observatory.

The different box sizes put different limits on the detection scales
in the simulations , which is $k \geq k_{box}=2\pi/L_{box}$. For our
simulations here, we have $k_{box}=0.0035, 0.0052, 0.01\mpchii$ for
$L1800$, $L1200$, and $L600$, respectively. Thus, in principle, we
expect to determine the behavior of the dark matter bispectrum on
scales larger than $k=0.1\mpchii$ with good precision. The large
resolution range used here enables us to check the consistency of
the results among different simulation boxes and to investigate the
behavior of dark matter bispectrum down to small scales. Here, we
choose the bin scheme as $\Delta log_{10}(k)=0.1$ for $k<0.1\mpchii$
and $0.05$ for the larger $k$ in Fourier space. But even for
$L1800$, we only have four realizations, which may still be not
enough to recover the exact variance of the bispectrum on large
scales. So, we calculate the uncertainty of bispectrum under the
assumption of a Gaussian field as (e.g.,
\citealt{Fry93,Scoccimarro98,Scoccimarro04,Sefusatti07})

\begin{eqnarray}
\Delta B^2_{123} &=& \frac{1}{N_{123}}\nonumber
P_{tot}(k_1)P_{tot}(k_2)P_{tot}(k_3) \\
\label{eqn:au} P_{tot}(k) &=& P(k)+\frac{1}{N_p},
\end{eqnarray}
where $N_{123}$ is the number of independent triangle configuration
modes in the Fourier space, and $P_{tot}(k)$ is the power spectrum
that includes the contribution of shot noise with $N_p$ particles.
Compared with Equation (28) in \cite{Sefusatti07}, there is no
$s_{123}$ factor in our Equation (\ref{eqn:au}) since $N_{123}$ is
the number of independent triangle modes and the statistically
rotational-invariance effect has already been taken into account.

\begin{figure}
\epsscale{1.2} \plotone{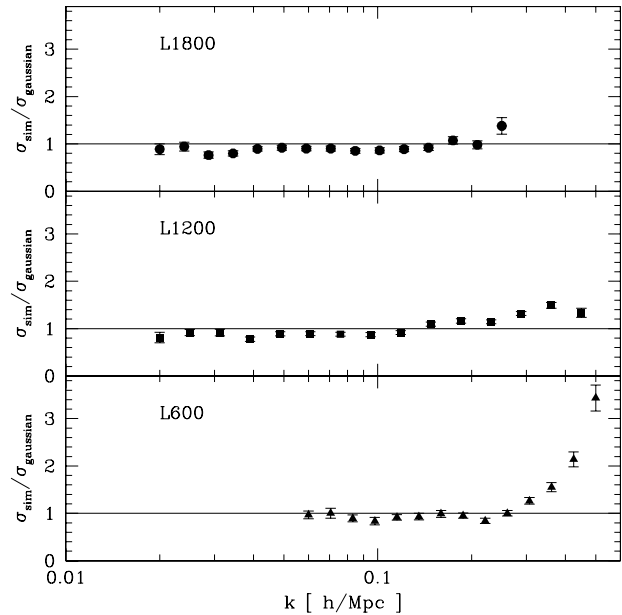} \caption{Ratios between the
variances of the mean of the reduced bispectrum averaged over the
realizations and the theoretical Gaussian uncertainties. Different
panels stand for different simulation boxes. The scale $k$ is
defined to be the longest side of a triangle.} \label{fig:sigr}
\end{figure}

We show the ratios between the variances of the mean of the reduced
bispectrum averaged over the realizations and the corresponding
uncertainties of a Gaussian field (Equation (\ref{eqn:au})) in
Figure \ref{fig:sigr}, where different panels stand for different
simulation boxes. The scale $k$ shown in Figure \ref{fig:sigr} is
defined to be the longest side of the triangle. In addition, to make
a complete statistics of the variance ratio, we have included all
the triangle configurations in each $k$ bin. On large scales of $k
\lesssim 0.2\mpchii$, the Gaussian uncertainties are slightly larger
than those determined from different simulation realizations, with
$\sigma_{sim}/\sigma_{gaussian}\thickapprox 0.9$. This is probably
due to the fact that we have used $P(k)$ determined from the
simulations to normalize the bispectra (see Equation
(\ref{eqn:qbp})). As the scale goes down to the nonlinear scales
where the assumption of a Gaussian field is not applicable, the
ratio increases rapidly, as shown in the case of $L600$. So, for our
purposes here, the utilization of Gaussian variance on the large
scales can, to some extent, compensate for still limited number of
independent realizations of our simulations.

\section{Results}
\label{results}

\subsection{Dark Matter Bispectrum}
\label{pt}

\begin{figure}
\epsscale{1.2} \plotone{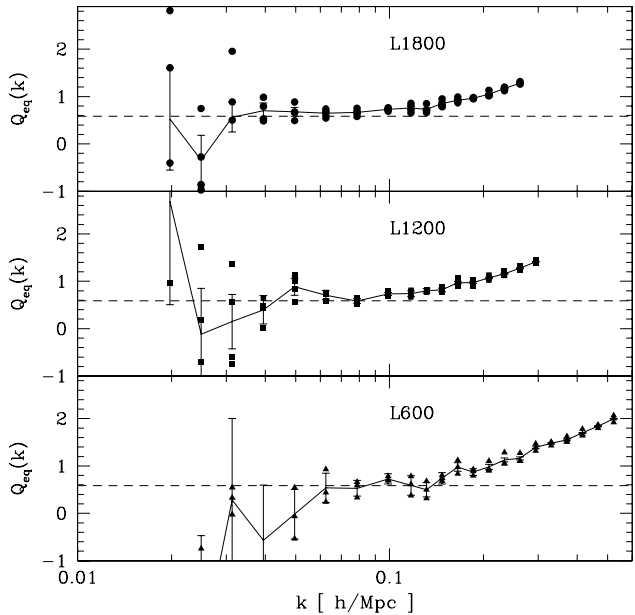} \caption{The scale dependence of
$Q_{eq}$. The points in different panels represent the different
realizations of each simulation set from $L1800$ to $L600$. The
solid lines with the error bars are the mean values with variances.
The variances here are obtained from the Gaussian uncertainties
(Equation\ref{eqn:au}). Also PT2 predictions are shown as the dashed
lines.} \label{fig:qeq}
\end{figure}

We mainly present our results with the reduced bispectrum
$Q(k,u,\alpha)$. Before we investigate in detail the scale and shape
dependence of $Q$, we check the consistency between PT2 and the
simulations for the equilateral triangle configurations, which are
defined that $k_1, k_2, k_3$ are in the same k-space bins. We show
the scale dependence of $Q_{eq}$ in Figure \ref{fig:qeq} where the
points in different panels represent the different realizations of
each simulation set from $L1800$ to $L600$. The solid lines with the
error bars are the mean values with variances. Here, the variances
of the mean are obtained from the Gaussian uncertainties instead of
the variances from different realizations in order to obtain an
accurate error estimation on large scales. Although they would be
underestimated on smaller scales, it does not affect our discussion
because we mainly focus on the large-scale behaviors. The PT2
predictions are shown as the dashed lines, with
$Q_{eq}=1/4+3\mu/4=0.585$ for our cosmological parameters.

From Figure \ref{fig:qeq}, we note that different $L_{box}$
simulations are consistent with each other on nonlinear scales of
$k>0.2\mpchii$ where $Q_{eq}$ deviates rapidly from the PT2
prediction as expected. But the situation is more complicated on
large scales where large fluctuations from realization to
realization are displayed, which is also implied by the Gaussian
uncertainties shown in the figure. For the case of $L1800$, which
clearly shows the trend of being constant at $k<0.1\mpchii$, the
mean of $Q_{eq}$ is, however, still about $20\%$ larger than that
predicted by PT2 at $0.04\mpchii<k<0.1\mpchii$. We also find such
deviations in $L1200$ at scales of $0.05\mpchii<k<0.1\mpchii$. For
$L600$ the tendency toward constant starts from $k \sim
0.06\mpchii$, but the large fluctuation prevents us from telling
whether the simulation has a similar deviation from the PT2. The
jump in the errors in $L600$ at $0.1\mpchii$ is caused by our change
of Fourier space bin scheme at this scale.

The fluctuation of $Q$ among the different simulation realizations
on large scales is probably due to the finite-volume effect or the
scarcity of realizations for each $L_{box}$. From the fluctuation,
we can determine the scales where enough independent triangle
configurations can be constructed to get an accurate value of $Q$.
As $L_{box}$ changes from $L1800$ to $L600$, the scales where $Q$
can be determined are $k >0.03 \mpchii$, $k >0.04 \mpchii$, and $k>
0.08\mpchii$, respectively if we require the errors are less than
about 30\%. Thus the finite-volume effect is still very important at
large scales of $k<0.1\mpchii$ even for simulations with a size as
large as that of $L600$. It would be extremely cautious to calculate
the dark matter bispectrum on scales of $k<0.1\mpchii$ when using
simulations with a size much smaller than that of $L600$. Moreover,
to make an accurate measurement on scales up to $k \sim
0.01\mpchii$, we still need to involve more realizations or
simulations of an even larger box size.

\begin{figure*}[t]
\epsscale{0.8} \plotone{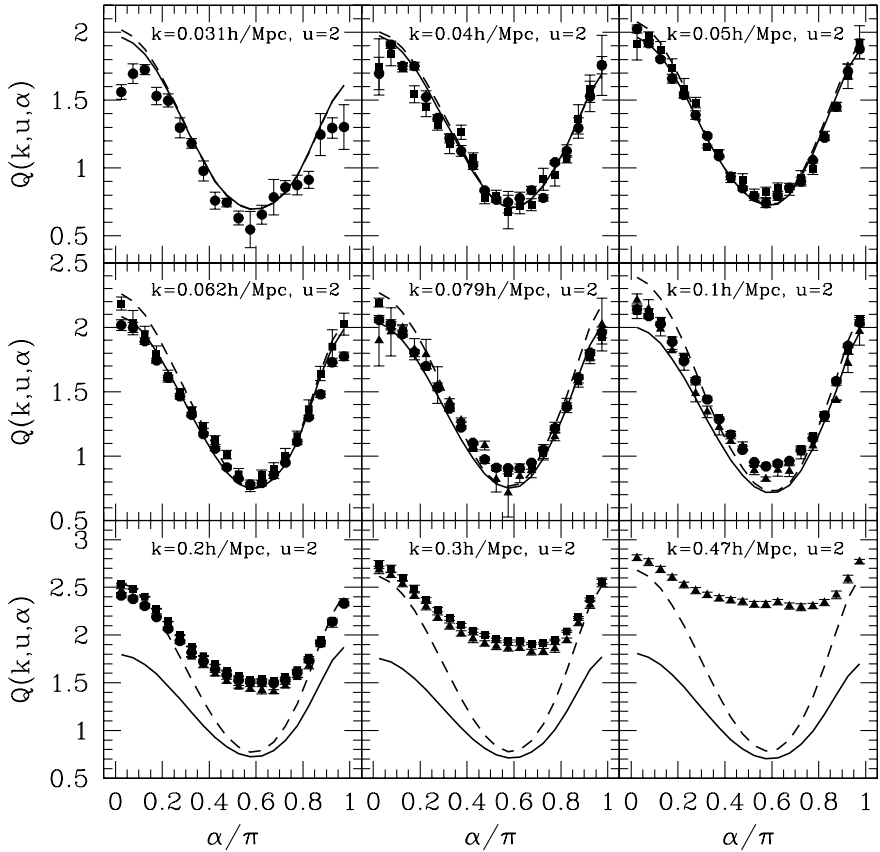} \caption{Scale and shape dependence
of $Q(k,u,\alpha)$ from large to small scales with $u\equiv2$, i.e.,
$k_2=2k_1$. Different points denote the results for different
$L_{box}$, with circles for $L1800$, squares for $L1200$, and
triangles for $L600$. The error bars are the variances on the
average of the different realizations. The dashed and solid lines
represent the PT2 predictions with the linear power spectrum
$P_L(k)$ and the nonlinear power spectrum $P_{NL}(k)$,
respectively.} \label{fig:qm}
\end{figure*}

In Figure \ref{fig:qm}, we show the scale and shape dependence of
$Q(k,u,\alpha)$ from large to small scales with $u \equiv
k_2/k_1=2$. Different points denote different $L_{box}$ results,
with circles for $L1800$, squares for $L1200$, and triangles for
$L600$. The error bars are the variances of the mean from the
different realizations. The dashed lines represent the PT2
predictions. As we see from the figure, different $L_{box}$
simulations are consistent with each other over all the scales. So,
different simulation resolutions do not affect the measurement of
$Q$ over all the scales here. The PT2 predictions only agree well
with the simulations on scales of $k<0.05\mpchii$ (where the
measurement errors are also large). Starting from $k \sim
0.06\mpchii$, PT2 has already deviated from the simulation
predictions, which indicates that the nonlinear effect may be
important even on these large scales.

We check the influence of nonlinear corrections by simply replacing
the linear power spectrum $P_{L}(k)$ in Equation (\ref{eqn:Bk}) and
Equation (\ref{eqn:qbp}) with the nonlinear power spectrum
$P_{NL}(k)$ derived directly from the simulations. The reason to do
this is that in real observations, both the bispectrum and the power
spectrum are nonlinearly evolved quantities.  The results are shown
as the solid lines in Figure \ref{fig:qm}. This nonlinear correction
can improve somewhat the match between the simulations and the PT2
at scales between $0.05\mpchii$ and $0.1\mpchii$, but does not
remove the mismatch completely.

Despite the overall agreement of PT2 with the simulations on large
scales for many different $\alpha$ configurations, PT2 still
overestimates the true $Q(k,u,\alpha)$ for coliner triangle
configurations, as shown clearly in the top left panel of Figure
\ref{fig:qm}. This phenomenon was also discovered by many other
authors(e.g., \citealt{Scoccimarro99,Smith08}). It may be explained
by the fact that these large-scale structures tend to be filamentary
rather than spherical in the simulations. For configurations near
that of the isosceles triangles where $k_3=k_2=2k_1$ and
$\alpha/\pi=0.58$, there is, however, a trend that PT2
underestimates the bispectrum from $k=0.06\mpchii$ to $0.1\mpchii$.
Such underestimates are not due to the shot-noise effect in the
simulations. We have corrected for the shot-noise effect when
obtaining $Q$. Because we have $1024^3$ dark matter particles for
each simulation, the shot noise is negligible for the final results.
So, such an inconsistency demonstrates the defect of PT2 on these
large scales. In the framework of perturbation theory, to describe
the simulations better, we might have to incorporate higher-order
corrections, such as the one-loop
correction(\citealt{Scoccimarro96,Bernardeau02}). Involving such
corrections may change the shape of $Q$ even on large scales, i.e.,
to make $Q$ smaller for colinear triangles and larger near isosceles
triangles, consistent with our results (e.g.,
\citealt{Bernardeau08}).

\subsection{The Halo Model}
\label{hm}
\subsubsection{Formalism}

To better understand the nonlinear effects in the reduced
bispectrum, we use the halo model method to theoretically study the
properties of $Q$(
\citealt{jmb98,Jing98,Ma00,Peacock00,Seljak00,Scoccimarro01,Berlind02,
Cooray02,Takada03,Wang04,Smith08}). The essence of the halo model is
to construct the distribution of particles(dark matter or galaxies)
by fixing the distribution of the clumps(dark matter halos) and the
distribution of the particles within the clumps on the assumption
that all the particles reside in the clumps. So, it is a halo-based
method and the changes in the properties of halos will affect the
final distribution of the particles.

Such a hierarchical method is very easy to apply given the
information of the different model components. The dark matter halo
distribution is generally constrained by the halo mass function
$n(m)$, which describes the dark matter halo number density $n$ as a
function of halo mass $m$. The particle distribution within the
halos is defined by the dark matter halo density profile
$\rho(r|m)$. So, the particle--particle correlations can be obtained
from the halo--halo correlations. By assuming a biased distribution
of halos relative to the underlying mass, we can describe the
halo--halo correlations with the halo bias parameters
$b_i(m)(i=1,2,\ldots)$\citep{Mo97,Sheth99}.

Following the notation described by \cite{Cooray02}, the dark matter
power spectrum can be decomposed into the 1-halo and 2-halo terms as

\begin{equation}
P(k) = P_{1h}(k)+P_{2h}(k) .
\end{equation}

These terms are

\begin{eqnarray}
P_{1h}(k) &=& M_{02}(k,k)  \nonumber \\
P_{2h}(k) &=& \left[ M_{11}(k)\right]^2 P_L(k) ,
\end{eqnarray}
where

\begin{eqnarray}
M_{ij}(k_1,\ldots,k_j) &\equiv& \int dm n(m)
\left(\frac{m}{\bar{\rho}}\right)^j b_i(m) \nonumber\\
&& \times [u(k_1|m) \ldots u(k_j|m)].
\end{eqnarray}
and $b_0\equiv 1$.

Here, $\bar{\rho}$ is the mean density of the universe and $u(k|m)$
is the normalized Fourier transform of the dark matter halo density
profile $\rho(r|m)$ truncated at the virial radius $r_{vir}$,

\begin{equation}
u(k|m) = \int_0^{r_{vir}} dr\ 4\pi r^2\,{\sin kr\over kr}\
{\rho(r|m)\over m}.
\end{equation}
As $k \rightarrow 0$, $u(k|m) \rightarrow 1$ and by definition, we
have

\begin{equation}
M_{11}(k) \rightarrow \int dm n(m)\frac{m}{\bar{\rho}} b_1(m)=1,
\label{eqn:m11}
\end{equation}
and so on large scales $P_{2h}(k) \rightarrow P_L(k)$, as expected.

The bispectrum is similarly decomposed into the contributions from
the 1-halo, 2-halo, and 3-halo terms,

\begin{equation}
B_{123} = B_{1h}+B_{2h}+B_{3h} ,
\end{equation}
where

\begin{eqnarray}
B_{1h} &=& M_{03}( k_1, k_2, k_3) , \nonumber \\
B_{2h} &=& M_{11}(k_1)M_{12}(k_2, k_3)
P_L(k_1) + {\rm cyc.} , \nonumber \\
B_{3h} &=& M_{11}(k_1)M_{11}(k_2)M_{11}(k_3)B_{PT}\nonumber \\
&& + \left[ M_{11}(k_1)M_{11}(k_2)M_{21}(k_3)P_L(k_1)P_L(k_2) + {\rm
cyc.} \right] . \nonumber \\
\label{eqn:Bk_th}
\end{eqnarray}

Again we have $B_{3h} \rightarrow B_{PT}$ as $k \rightarrow 0$,
which means that the 3-halo term will return to that of the PT2
predictions on large scales.

The reduced bispectrum $Q$ is then defined as

\begin{equation}
Q(k_1,k_2,k_3)=\frac{B_{123}}{P(k_1)P(k_2)+cyc}.\label{eqn:qbp_th}
\end{equation}
Note the difference from Equation (\ref{eqn:qbp}) in the denominator
where the linear power spectrum $P_L(k)$ is replaced by the
nonlinear power $P(k)$ determined in the halo model.

In our model, dark matter halos are defined as objects with a mean
density $\Delta_{vir}$ times that of the background
universe\citep{Bryan98}, where $\Delta_{vir} \approx 361$ for our
cosmology parameters, and their density distribution follow the
Navarro--Frenk--White (NFW) profile \citep{Navarro96}. The
concentration parameter $c(m)$ in the density profile is given by
the relation $c(m)=c_0(m/m_*)^{\beta}$, where $c_0=9$,
$\beta=-0.13$, and $m_*=4.8\times10^{12}\msun$ is the nonlinear mass
scale\citep{Bullock01}. We also generate the linear power spectrum
from CMBfast here.

For the halo mass function(MF), first we consider the two commonly
used analytical models of \cite{Press74}(PS) and \cite{Sheth99}(ST).
They are defined by

\begin{equation}
n(m)dm = \frac{\bar{\rho}}{m}f(\nu)d\nu,
\end{equation}
where $\nu=\delta_c/\sigma(m)$, with $\delta_c=1.69$, the spherical
collapse threshold of the linear overdensity $\delta$. $\sigma(m)$
is the linear rms fluctuation within spheres of radius $R =
(3m/4\pi\bar{\rho})^{1/3}$,

\begin{equation}
 \sigma^2(m) =
     \int \frac{{\rm d}k}{k} {k^3P_L(k)\over
2\pi^2}W^2(kR),\label{eqn:sigma}
\end{equation}
where $W(x)=3(\sin x-x\cos x)/x^3$ is the Fourier transform of the
top-hat window function and $W(x) \rightarrow 1$ as $x \rightarrow
0$.

The function $f(\nu)$ for PS and ST can be generalized into the
following form:

\begin{equation}
\nu f(\nu)=2A\sqrt{\frac{a\nu^2}{2\pi}}\left[1+(a\nu^2)^{-p}\right]
\exp\left(-\frac{a\nu^2}{2}\right)
\end{equation}
where $A=1, 0.322$, $p=0, 0.3$, and $a=1, 0.707$ for PS and ST,
respectively.

\cite{Jenkins01} have also derived the halo mass functions for
different halo definitions of FOF halos and spherical
overdensity(SO) halos. Here, we only use their FOF halo mass
function to match our halo definition

\begin{equation}
\nu f(\nu) = 0.307\exp(-|\ln(\nu/\delta_c)+0.61|^{3.82}),
\end{equation}
with $-0.9 \leq \ln(\nu/\delta_c) \leq 1.0$, corresponding to about
$4\times10^{11}\msun \leq m \leq 4\times10^{15}\msun$ in our
simulations. Although this mass function is derived within a finite
halo mass range and does not satisfy the density normalization
requirement, we can still impose the normalization constraint of
Equation (\ref{eqn:m11}) by only integrating over a limited mass
range, i.e.,
\begin{eqnarray}
\int_0^{\infty} dm n(m)\frac{m}{\bar{\rho}} b_1(m)
=\int_{M_1}^{M_2} dm n(m)\frac{m}{\bar{\rho}} b_1(m) + \nonumber \\
\left[1-\int_{M_1}^{M_2} dm n(m)\frac{m}{\bar{\rho}} b_1(m)\right].\nonumber\\
\label{eqn:normalization}
\end{eqnarray}
The upper mass limit $M_2$ is usually set by the maximum halo mass
in the simulations. Here, we just use the maximum halo mass seen in
$L1800$ as $M_2=6\times10^{15}\msun$.

For the halo bias parameters, we use the results of \cite{Mo97} and
\cite{Sheth99}. We only need to consider the first two halo bias
parameters:

\begin{eqnarray}
b_1(m) &=& 1 + \epsilon_1 + E_1, \\
b_2(m) &=& 2(1+a_2)(\epsilon_1 + E_1) + \epsilon_2 + E_2,
\end{eqnarray}
where

\begin{eqnarray}
&&\epsilon_1 = \frac{a\nu^2 - 1}{\delta_c},\ \
\epsilon_2 = \frac{a\nu^2}{\delta_c^2}(a\nu^2-3), \nonumber \\
&&E_1 = \frac{2p/\delta_c}{1 + (a\nu^2)^p},\ \  \frac{E_2}{E_1} =
\frac{1 + 2p}{\delta_c} + 2\epsilon_1, \nonumber
\end{eqnarray}
\begin{equation}
{\rm and}\ \ \ \ \ a_2 = -17/21. \nonumber
\end{equation}
As for the \cite{Jenkins01} MF(hereafter JMF), we also use the bias
parameters corresponding to the PS MF. Changing to the other halo
bias model does not significantly affect the predictions for the
power spectrum or bispectrum.

\begin{figure}
\epsscale{1.2} \plotone{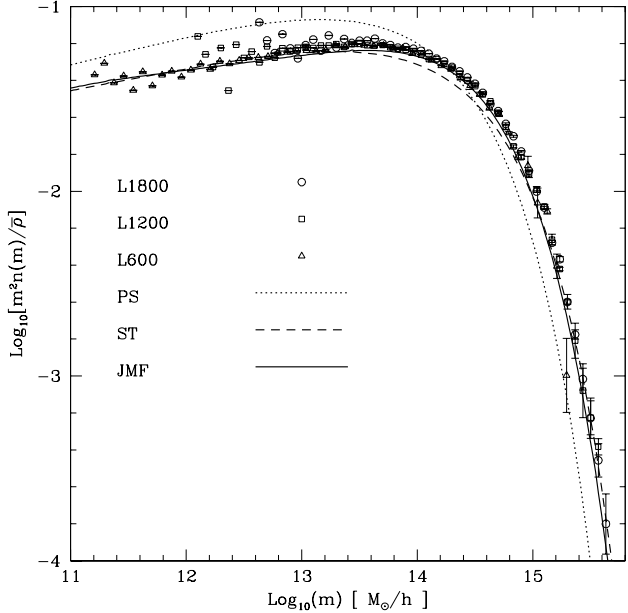} \caption{Three mass functions
compared with our simulations. Different points stand for different
$L_{box}$ simulations and different lines for different mass
functions} \label{fig:mfun}
\end{figure}

We compare the three mass functions with our simulations for
$m^2n(m)/\bar{\rho}$ in Figure \ref{fig:mfun}, where different
points stand for different $L_{box}$ simulations and different lines
for different mass functions. The error bars are determined from the
scatter among the different realizations. The fluctuations shown in
the low mass ends of different $L_{box}$ simulations are caused by
our lowest mass halo definitions that we set $10$ particles as the
minimum requirement for a bound halo. Except for the PS MF, the
other MFs have similar halo number densities at the high mass end,
though JMF is somewhat smaller than ST MF for $m>10^{15}\msun$. The
figure indicates that PS MF has much fewer large halos while
retaining too many small mass halos. The ST MF is consistent with
the simulations for the high mass and low mass halos, but it is
substantially lower in the range of $10^{13}\msun$--$10^{15}\msun$,
which is quite important for the scales we consider here. JMF seems
to agree with the simulations better for these intermediate masses,
though in a detailed comparison we find that it is still lower than
the simulations for $m>10^{13}\msun$.

In our simulations, about $20\%$--$30\%$ of the total halo particles
reside in halos with $m \geq 10^{14}\msun$ and about $53\%$--$80\%$
are in halos of $m \geq 10^{13}\msun$. The fluctuations are due to
the different resolutions of different $L_{box}$ simulations.
Therefore, the differences among different MFs in the mass range of
$10^{13}\msun$--$10^{15}\msun$ will strongly affect the final
results at least in the intermediate scales as shown in Figure
\ref{fig:pnl}. Although we do not have a mass function exactly
consistent with the simulations, we can still compare the results of
$Q$ for these three MFs to study the effect of different halo mass
distribution.

\subsubsection{The Halo Model Results}

\begin{figure}
\epsscale{1.2} \plotone{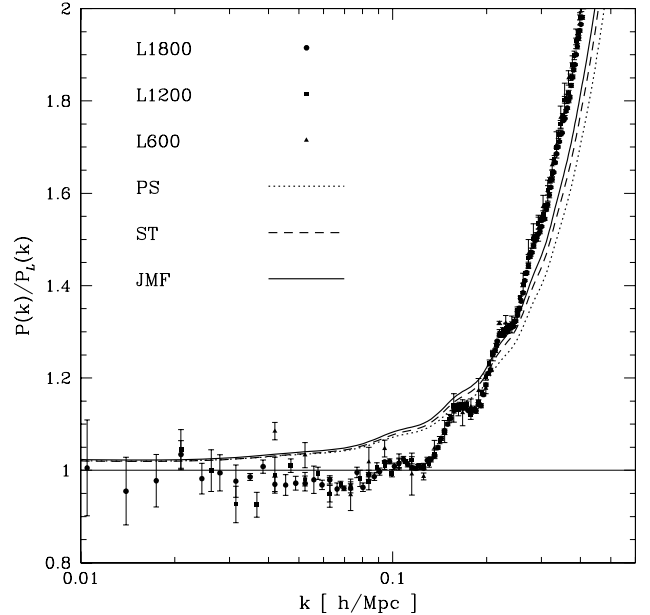} \caption{Ratio between the nonlinear
power spectrum $P(k)$ and the linear power $P_L(k)$ for the halo
model predictions with different MFs, compared with those of the
simulations.} \label{fig:pnl}
\end{figure}

After specifying the MF and the halo bias, we can theoretically
determine the power spectrum and the bispectrum of the dark matter.
In Figure \ref{fig:pnl}, we show the ratio between the nonlinear
power spectrum $P(k)$ and the linear power $P_L(k)$ for halo models
of different MFs, compared with those of the simulations. The
wiggles on the large scales are caused by the fact that the Baryonic
Acoustic Oscillations(BAO) features are suppressed in the nonlinear
evolution relative to the linear evolution. The nonlinear power
spectrum from the simulations is a bit smaller than the linear power
on large scales (corresponding to the first BAO peak), and it
rapidly increases when $k>0.1\mpchii$. The halo models for different
MFs converge on large scales where the 2-halo term dominates. But
they are still larger than the linear power spectrum, because the
1-halo term still has a nonvanishing contribution to $P(k)$ on these
large scales. Compared with the simulations, the power spectra of
the halo models are larger than those of the simulations, and the
wiggles in $k>0.1\mpchii$ where the 1-halo term becomes important
are not as evident as in the simulations. For scales of
$k>0.2\mpchii$, all the halo model predictions are substantially
lower than those of the simulations.  On such scales, the large mass
halos play an important role in the 1-halo term. And yet we have not
corrected for the halo boundary and exclusion effects(e.g.,
\citealt{Takada03}) in the halo models. The effect of choosing an MF
is relatively small, as the predicted power spectra based on the
three MFs are very similar. The simple analytical prescription of
the halo model, instead of its ingredients (the MF, the halo density
profile, and the halo bias factors), may be the main cause leading
to the discrepancies between the models and the simulations shown in
Figure \ref{fig:pnl}.

\begin{figure*}[t]
\epsscale{0.8} \plotone{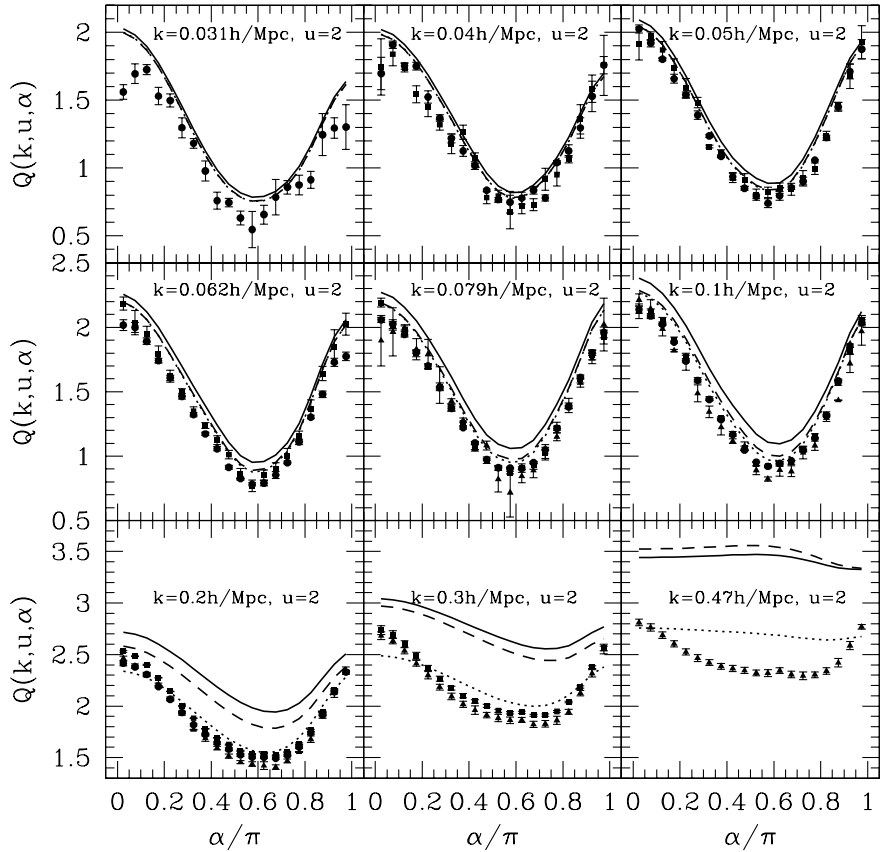} \caption{Same as Figure
\ref{fig:qm}, but for the halo model predictions with different MFs.
The points are for the simulation results. The dotted, dashed, and
solid lines show the halo model predictions with PS, ST, JMFs
respectively.} \label{fig:qm_th}
\end{figure*}

In Figure \ref{fig:qm_th}, we show the halo model predictions for
the reduced bispectrum $Q(k,u,\alpha)$. As in Figure \ref{fig:qm},
the points are for the simulations. The dotted, dashed, and solid
lines show the halo model predictions for PS, ST, JMFs respectively.
Compared with the PT2 predictions in Figure \ref{fig:qm}, the halo
models are a bit larger on large scales, as shown in the top panels
of Figure \ref{fig:qm_th}. Different halo model predictions converge
on the large scales where the 3-halo term dominates. But the
predictions of the JMF model are somewhat larger than the others. It
is generated by the fact that the 2-halo term contribution is much
higher for JMF. Such differences are shown clearly in the
intermediate scales of $0.06$--$0.1\mpchii$ in the middle panels of
Figure \ref{fig:qm_th}. On these scales, the halo models capture the
shape of reduced bispectrum in the simulations rather well compared
with the PT2 predictions, though the amplitudes are still higher
than expected. The result for the PS MF seems to fit with the
simulation bispectrum better near the isosceles triangle shapes. For
the quasi-linear and nonlinear scales of $0.2$--$0.5\mpchii$ in the
bottom panels of Figure \ref{fig:qm_th}, the differences between the
MFs show up distinctly. The model predictions for the PS MF are in
much better agreement with the simulations. Even at the highly
nonlinear scale of $k \sim 0.5\mpchii$, it can provide a fairly good
estimate of $Q$. And as $k$ increases, the reduced bispectra of the
simulations approach a constant more slowly than the halo models
predict. Moreover, the turnover feature in the shape dependence,
appeared in the halo model predictions for the ST and JMFs at the
scale of $k \sim 0.5\mpchii$, is not seen in the simulations.

The better agreement on the reduced bispectrum between the halo
model predictions for the PS MF and the simulations on scales of
$k>0.2\mpchii$ does not mean that PS MF is a better description for
the mass function in halo models. Since the reduced bispectrum is
the ratio of bispectrum to a sum of power spectrum products, the
better agreement achieved with the PS MF is largely a coincidence,
because the power spectra are fitted worse on relevant scales by the
PS model(as shown in our Figure \ref{fig:pnl}). In fact, the ST MF
and JMF better describe the mass function of halos in our
simulations (Figure \ref{fig:mfun}). We also note that our result is
consistent with \citet{Fosalba05}, who argued that a better
agreement in the reduced bispectrum between the halo model and
simulations can be achieved by imposing a mass cutoff for halos at
the high mass end. But, we do not think that imposing the mass
cutoff is physical in our study, since our simulation box is big
enough that the halos at high mass end of ST MF (or JMF) are fairly
well sampled in our simulations. Therefore, the failure of the halo
model for the ST MF or JMF implies the inherent difficulty of the
halo models to accurately predict the power spectrum and the
bispectrum.

\begin{figure}
\epsscale{1.2} \plotone{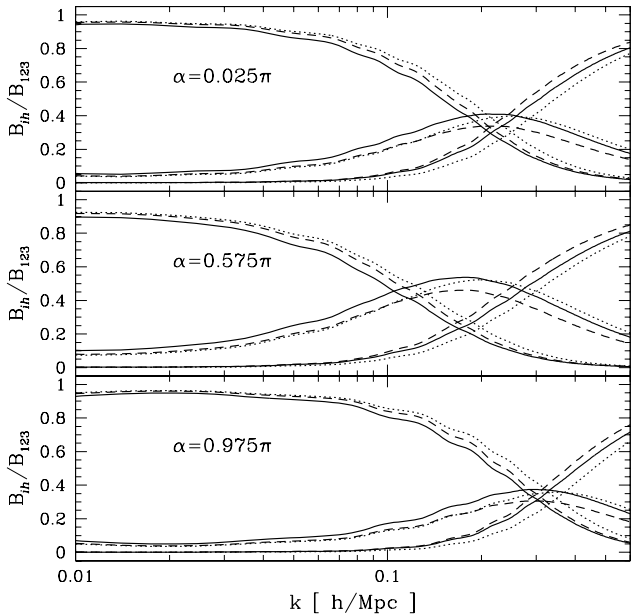} \caption{Scale and shape dependence
of each halo term contribution $B_{\it{i}h}/B_{123}$ with $\mu
\equiv k_2/k_1 = 2$, as well as the effects of the different MFs. We
show from top to bottom for three triangle shapes including the
colinear and isosceles triangles, specified by different $\alpha$
values. At scale $k=0.01\mpchii$, in each panel from top to bottom
are the 3-halo, 2-halo, and 1-halo term contributions. Different
lines denote the results for different MFs as in Figure
\ref{fig:pnl} and Figure \ref{fig:qm_th}.}\label{fig:bcon}
\end{figure}

To investigate the effects of choosing different MFs and the
contributions of different halo terms, we show in Figure
\ref{fig:bcon} the scale and shape dependence of
$B_{\it{i}h}/B_{123}$ of each halo term by changing $k$ and $\alpha$
while fixing $\mu \equiv k_2/k_1 = 2$. We show from top to bottom
for three triangle shapes including the colinear and isosceles
triangles, specified by different $\alpha$ values. At the scale
$k=0.01\mpchii$, in each panel from top to bottom are the 3-halo,
2-halo, and 1-halo term contributions. Different lines denote the
halo model predictions for different MFs as in Figures \ref{fig:pnl}
and \ref{fig:qm_th}. On large scales of $k<0.1\mpchii$, the 3-halo
term dominates but 2-halo term still has a contribution of about
$5\%$--$20\%$ in these halo models. Comparing Figure \ref{fig:qm}
and Figure \ref{fig:qm_th}, it seems that the 2-halo term
contributions in the halo models are higher than simulations
predict. Furthermore, we note that in the determination of the
reduced bispectrum $Q(k,u,\alpha)$ of the halo models, the linear
power spectrum $P_L(k)$ in the denominator of Equation
(\ref{eqn:qbp}) in PT2 is replaced by the nonlinear power $P(k)$
which is larger than $P_L(k)$ over all the scales shown in Figure
\ref{fig:pnl}. Therefore, the 2-halo term is actually necessary to
fit with the simulations, otherwise only using the 3-halo
term($B_{3h} \rightarrow B_{PT}$ on large scales) will make
$Q(k,u,\alpha)$ of the halo models much smaller than those of
simulations. For the different halo term contributions, the
differences among choosing different MFs are very small on large
scales. The PS and ST MFs have nearly the same halo-term
contributions at these scales. This indicates that the bispectrum on
these scales is not sensitive to the choice of MFs.

When $k>0.1\mpchii$, we see from Figure \ref{fig:bcon} that the
2-halo term is much more important for the isosceles triangle shapes
than for the colinear ones. It has almost the same contribution as
the 3-halo term at $k \sim 0.1\mpchii$ for isosceles shapes, while
the 3-halo term still dominates in the colinear configurations.
Thus, it explains the failure of PT2 which only includes the 3-halo
term for the isosceles triangles, shown in the middle panels of
Figure \ref{fig:qm}. On these quasi-linear scales, the 1-halo term
also starts to become important and its contribution differs for
different triangle shapes, which clearly shows the influence on the
shape dependence of $Q(k,u,\alpha)$. Again, the 1-halo term is most
important for the isosceles triangle configurations. We can
therefore infer that the nonlinear effect is more significant for
such triangle shapes. Meanwhile, the differences also emerge on
these quasi-linear scales when different MFs are used. We note from
Figure \ref{fig:qm_th} that the reduced bispectrum for the ST MF
case increases rapidly in the quasi-linear and nonlinear regime.
From Figure \ref{fig:bcon}, we can see that it is caused by the
rapid increase in the 1-halo term contribution.  Even in the
nonlinear scale of $k=0.5\mpchii$, where the 1-halo term is
dominant, the 2-halo contribution is still not negligible and the
3-halo term also has a nonvanishing contribution for some triangle
configurations.

Despite the fact that the halo model can provide a qualitative
explanation for $Q(k,u,\alpha)$ down to the quasi-linear scales,
there still exist unavoidable problems with the halo model if we
consider the possibility that the halo model can match the power
spectrum $P(k)$ of the simulations extremely well. We take
$k=0.1\mpchii$ for example. We see from Figure \ref{fig:pnl} that
the nonlinear power $P(k)$ of the simulations is almost equal to the
linear power $P_L(k)$, only with a small deviation at this scale.
Then we can replace $P(k)$ in Equation (\ref{eqn:qbp_th}) with
$P_L(k)$. Thus, the reduced bispectrum $Q(k,u,\alpha)$ must be
larger than $Q_{PT}$ in PT2, because the 3-halo term has already
included $B_{PT}$ (at this scale $M_{11}(k)$ is still very close to
1) but the 2-halo term can even add an additional sizable
contribution. For the case of colinear configurations with $\alpha
\rightarrow 0$, in Figure \ref{fig:qm}, the PT2 prediction is much
larger than those of the simulations. Therefore, the halo model
prediction with the 2-halo term included would give an even larger
bispectrum. This apparent contradiction in the predictions for
$P(k)$ and $Q(k,u,\alpha)$ between the halo models and the
simulations actually comes from the prescription of the halo model.
Because in the halo model, it includes the PT2 predictions in the
3-halo term in Equation (\ref{eqn:Bk_th}). The failure of PT2 will
then lead to the failure of the halo model if its predictions are
larger than the simulations. Moreover, the 2-halo term and even the
1-halo term are not negligible at relevant large scales.

\section{Discussions and Conclusions}
\label{conclusions}

We use a set of numerical $N$-body simulations to study the
large-scale behavior of the reduced bispectrum $Q(k,u,\alpha)$ and
compare it with the analytical predictions from both the
second-order perturbation theory (PT2) and the halo models. We
investigate the scale and shape dependence of $Q$ and find that
although PT2 agrees fairly well with the simulations on the large
scales of $k<0.05\mpchii$, it still shows a signature of deviation
as the scale goes down. Even on the largest scale where the
bispectrum can be explored with our simulations, the inconsistency
between PT2 and the simulations appears for the colinear triangle
shapes. The better agreement of PT2 with the simulations after the
nonlinear power spectrum is used for the PT2 implies the importance
of nonlinear effects even on these large scales of $k<0.1\mpchii$,
which is a supplement to the previous studies(e.g.,
\citealt{Bernardeau02,Smith08}). Then one of our main conclusions is
that higher-order corrections (e.g., loop corrections) are necessary
on large scales of $k= 0.05$--$0.1 \mpchii$ for the perturbation
theories to predict $Q$ at an accuracy of percent level.

An alternative method of predicting the reduced bispectrum is
through the halo model with several basic ingredients.  After a
detailed comparison between the predictions and our simulation
results, however, we are disappointed to find that halo models can
at best serve as a qualitative method to help study the behavior of
$Q$ on large scales and also on relatively small scales. But because
of the nonnegligible two-halo term contribution, the halo model
predictions of the reduced bispectrum are always higher than the
results of the simulations.  No halo model can be brought into
agreement with the bispectrum on nonlinear scales unless a
reasonable mass function is adopted.

We have carefully and accurately measured the bispectrum for a large
range of scales in the simulations. The results could be useful for
future analytical modeling of $Q$. Therefore, a table of $Q$ from
our measurement is made available at {\it
http://www.shao.ac.cn/mppg/guo/paper/dmbisp.html}.

\acknowledgments

We thank Jun Pan and Roman Scoccimarro for useful discussions.  This
work is supported by NSFC (10533030, 10821302, 10873028, 10878001),
by the Knowledge Innovation Program of CAS (no. KJCX2-YW-T05), and
by 973 Program (no.2007CB815402).

\end{document}